\documentclass[aps,twocolumn,showpacs,superscriptaddress,groupedaddress,10pt]{revtex4-1}
\usepackage{graphicx}
\usepackage{mathtools}
\usepackage[dvips]{color}
\usepackage{epsfig}
\usepackage{amssymb}
\usepackage{amsmath}
\usepackage{bm}
\usepackage{float}

\usepackage{varioref}

\begin{document}

\bibliographystyle{apsrev}
\title{Asymmetric Mutualism in Two- and Three-Dimensional Range Expansions}
\author{Maxim O. Lavrentovich}
\email{lavrentm@gmail.com}
\affiliation{Department of Physics, Harvard University, Cambridge, Massachusetts 02138, USA}
\author{David R. Nelson}
\email{nelson@physics.harvard.edu}
\affiliation{Department of Physics, Harvard University, Cambridge, Massachusetts 02138, USA}
\begin{abstract}
Genetic drift at the frontiers of two-dimensional range expansions of microorganisms can frustrate local cooperation between different genetic variants, demixing the population into distinct sectors. In a biological context, mutualistic or antagonistic interactions will typically be asymmetric between variants.  By taking into account both the   asymmetry and the  interaction strength, we show that the much weaker demixing in three dimensions allows for a mutualistic phase over a much wider range of asymmetric cooperative benefits, with mutualism prevailing for any positive, symmetric benefit. We also demonstrate that expansions with undulating fronts  roughen dramatically at the boundaries of the mutualistic phase, with severe consequences for the population genetics along the transition lines.
\end{abstract}

\date{\today}

 \maketitle


When a population colonizes new territory, the abundance of unexploited resources allows the descendants of the first few settlers to thrive.  These descendants invade the new territory and form genetically distinct regions or sectors at the population frontier. If the frontier population is small,  the birth and death of individuals create large fluctuations in the sector sizes.  These fluctuations, called genetic drift, cause some settler lineages to become extinct as neighboring sectors engulf their territory.  Over time, this sector coarsening process dramatically  decreases genetic diversity at the frontier \cite{KorolevRMP, MKNPRE}.

 Interactions between the organisms can modify the coarsening process.  For example, cooperative interactions, in which  genetic variants in close proximity confer growth benefits upon each other, can lead to the founders' progeny remaining intermingled.  Then, coarsening  does not occur, and the consequent growth pattern is called a  ``mutualistic phase'' \cite{KorolevMut}.    Cooperative interactions are commonly found in nature:  microbial strains exchange resources \cite{microbemut}, ants protect aphids in exchange for food \cite{insectmut}, and different species of mammals share territory to increase foraging efficiency \cite{mammalmut}.  Recently, a mutualistic phase was experimentally realized in partner yeast strains that feed each other \cite{MuellerMut, WashingtonMut}. These experiments require an understanding of   \textit{asymmetric} interactions  where species do not benefit equally from cooperation. Antagonistic interactions may also occur, e.g., between bacterial strains secreting antibiotics against competing strains \cite{antagonism}. These interactions and the mutualistic phase also play prominent roles in theories of nonequilibrium statistical dynamics   \cite{KorolevMut, Beghe, Hinrichsen, NPRGVoter, CardyTauber, hammal, dornic}.

We explore   here asymmetric cooperative and antagonistic interactions in two- and three-dimensional range expansions.  Two-dimensional expansions $(d=1+1)$ occur when the population grows in a thin layer, such as in a biofilm or on a Petri dish.  Three-dimensional expansions $(d=2+1)$ occur, for example, at the boundaries of growing avascular tumors \cite{cancer5, cancer3, cancer6}.  We model both flat and rough interfaces at the frontier, the latter being an important feature of many microbial expansions \cite{DRNPNAS}. 

We arrive at the following biologically relevant results:  Three-dimensional range expansions support mutualism more readily than planar ones,  and a mutualistic phase occurs for \textit{any}  symmetric cooperative benefit. Conversely, two-dimensional expansions require a critical  benefit \cite{KorolevMut,Beghe}.   In addition, we  find that the frontier roughness is strongly enhanced at the onset of mutualism for asymmetric interactions. Finally, we find that frontier roughness allows for a mutualistic phase over a wider range of cooperative benefits.

 \textit{Flat Front Models.}--- We consider two genetic variants, labelled black and white.  Cells  divide only at the population frontier \cite{KorolevRMP}.   Such expansions occur when nutrients are absorbed before they can diffuse into the interior of the population, inhibiting cell growth behind the population front.  This can occur in tumor growth \cite{cancer6} and in  microbial expansions at low nutrient concentrations \cite{shielding}.

\begin{figure}[!ht]
\includegraphics[height=2.65in]{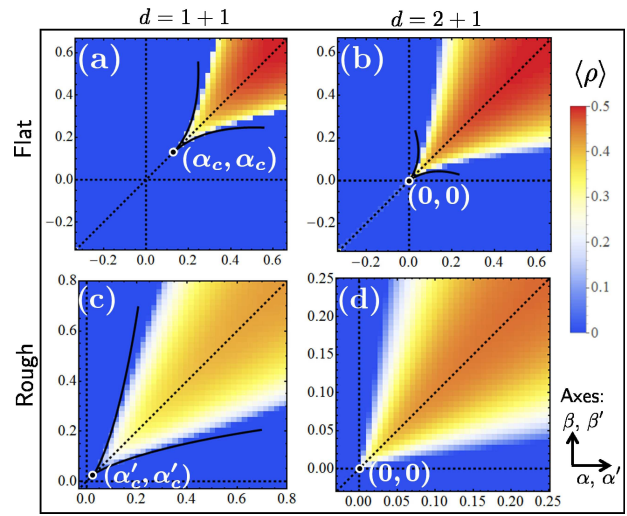}
\caption{\label{fig:PhaseDiagram}  (Color online) The density $\langle \rho \rangle$  of genetic sector interfaces along the population front at long times $t$ (measured in generations), averaged over $10^3$ runs of the flat [$(a)$  and $(b)$] and rough [$(c)$ and $(d)$]  front range expansion models for $d=1+1$ [$(a)$ and $(c)$] and $d=2+1$ [$(b)$ and $(d)$]  (system sizes $L$ and $L^2$ cells, respectively) for cooperative benefits $\alpha$ and $\beta$ defined in Eq.~(\ref{eq:updaterules}) [$\alpha'$ and $\beta'$ for rough fronts in Eq.~(\ref{eq:birthrates})].  The  parameters are: $(a)$  $t= 1.5\times10^5$, $L=8\times10^3$; $(b)$ $t=3\times10^3$, $L^2=600^2$;  $(c)$ $t= 2.5\times10^3$, $L=10^3$; $(d)$ $t = 500$, $L^2=50^2$;  There is a mutualistic phase in the $\alpha,\beta>0$ ($\alpha',\beta'>0$)  quadrant in all panels.  The solid lines in $(a)$, $(b)$ and $(c)$ show the directed percolation   (DP) transition line shapes near the bicritical points (see the Appendix for details). A symmetric DP (DP2) point occurs at $\alpha = \beta = \alpha_c = 0.1242(5)$   in $(a)$ and a  ``rough DP2'' point at $\alpha' = \beta' = \alpha_c' = 0.0277(2)$ in $(c)$.   The dotted lines indicate the loci $\alpha=0$, $\beta=0$, and $\alpha = \beta$.}
\end{figure}

  According to a continuum version of a stepping stone model  \cite{kimurapaper, KorolevMut}, the coarse-grained fraction of black cells $f \equiv f(\mathbf{x},t)$ at some position $\mathbf{x}$ along a \emph{flat} population front at time $t$ obeys
\begin{equation}
\partial_t f  = D \nabla^2 f+   \tau_g^{-1}  f(1-f) \left[s \left(  \frac{1}{2} -f\right) + \frac{r}{2  }   \right] +\eta,   \label{mutlang}
\end{equation}
where $D$ is a  diffusivity,  $\eta (\mathbf{x},t)$ an \^Ito noise term  \cite{KorolevRMP} with $\langle \eta(\mathbf{x},t) \eta(\mathbf{x}',t) \rangle= 2 \ell^{d_s} \tau_g^{-1}  f(1-f) \delta(\mathbf{x}-\mathbf{x}')\delta(t-t')$, $\tau_g$  a generation time,  $d_s$  the spatial dimension, and $\ell$  an effective lattice spacing.    Also, $r=\alpha-\beta$ and $s = \alpha+\beta$, where $\alpha$ and $\beta$ represent the increase in growth rates over the base rate per generation of the black and white species, respectively, in the presence of the other species.     Equation~(\ref{mutlang}) describes the behavior of  two-dimensional expansions of mutualistic strains of yeast \cite{MuellerMut} and is expected to characterize many different range expansions \cite{KorolevRMP,KorolevMut}. At $r=s=0$, Eq.~(\ref{mutlang}) reduces to the Langevin equation of the voter model  \cite{dornic, hammal}. A signature of the mutualistic phase is a non-zero average density  $\langle \rho \rangle$ of black and white cell neighbors (genetic sector interfaces) at the front at long times.  To construct phase diagrams for our different models in Fig.~\ref{fig:PhaseDiagram}, we calculate $\langle \rho \rangle$ as a function of the cooperative benefits $\alpha,\beta \geq 0$  and antagonistic interactions $\alpha,\beta < 0$.

\begin{figure}[ht]
\includegraphics[height=2.8in]{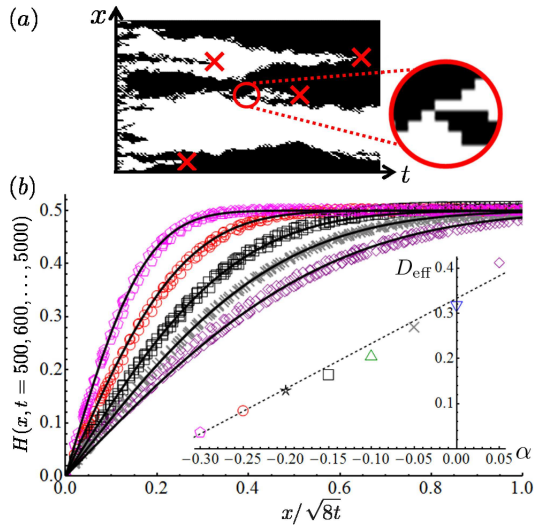}
\caption{\label{fig:FlatHet}  (Color online)   (a)   A sample evolution of the flat front model for $d=1+1$  at $\alpha = \beta = 0$ (front size $L=100$).   The rules in Eq.~(\ref{eq:updaterules}) allow interface branching (circled event),  facilitating mutualistic mixing. The coarse-grained dynamics for $\alpha=\beta<\alpha_c$ are dominated by  pair annihilations of  interfaces (marked by crosses)   \cite{NPRGVoter, CardyTauber}. (b)  We collapse the heterozygosity $H(x,t)$,  averaged over  1600  runs with $L=10^3$, at different times (points in main plot)  and fit to Eq.~(\ref{hetcollapse})  (solid lines) to find $D_{\mathrm{eff}}$ in the inset.  The dashed line  indicates the prediction   $D_{\mathrm{eff}} = 1/3+\alpha$. This prediction is discussed in the Appendix. }
\end{figure}

We propose a microscopic model for flat fronts in the spirit of Grassberger's  cellular automaton \cite{Grassberger1}, which obeys Eq.~(\ref{mutlang}) under an appropriate coarse-graining, as we verify in the following.  Domain wall branching, shown in Fig.~\ref{fig:FlatHet}$(a)$, is required for a mutualistic phase in $d=1+1$ dimensions.   Our model  allows branching by having three cells compete to divide into a new spot on the frontier during each time step.  To approximate cell rearrangement at the frontier, we assume that the order of competing cells in each triplet is irrelevant.  The update rules are:
\begin{equation}
\begin{cases}
p\bigg( \begin{tabular}{c}  $\circ \bullet \circ$ \\[-6pt] ${\scriptstyle \Downarrow}$ \\[-5pt] $\bullet$ \end{tabular}\bigg) =p\bigg( \begin{tabular}{c}  $\bullet \circ \circ$ \\[-6pt] ${\scriptstyle \Downarrow}$ \\[-5pt] $\bullet$ \end{tabular}\bigg) =p\bigg( \begin{tabular}{c}  $\circ \circ \bullet$ \\[-6pt] ${\scriptstyle \Downarrow}$ \\[-5pt] $\bullet$ \end{tabular}\bigg) = \dfrac{1}{3} +\alpha \\[-4pt]
p\bigg( \begin{tabular}{c} $\bullet \circ \bullet$ \\[-6pt] ${\scriptstyle \Downarrow}$ \\[-5pt] $\circ$ \end{tabular}\bigg) =p\bigg( \begin{tabular}{c} $\circ \bullet \bullet$ \\[-6pt] ${\scriptstyle \Downarrow}$ \\[-5pt] $\circ$ \end{tabular}\bigg) =p\bigg( \begin{tabular}{c} $\bullet \bullet \circ$ \\[-6pt] ${\scriptstyle \Downarrow}$ \\[-5pt] $\circ$ \end{tabular}\bigg) = \dfrac{1}{3}+\beta \\[-4pt]
p\bigg( \begin{tabular}{c} $\bullet \bullet \bullet$ \\[-6pt] ${\scriptstyle \Downarrow}$ \\[-5pt] $\bullet$ \end{tabular}\bigg) = p \bigg( \begin{tabular}{c} $\circ \circ \circ$ \\[-6pt] ${\scriptstyle \Downarrow}$ \\[-5pt] $\circ$ \end{tabular}\bigg) =1, 
\end{cases} \label{eq:updaterules}
\end{equation}  

 where $-1/2 \leq \alpha,\beta \leq 2/3$.     The rules for all other combinations follow from probability conservation.

Positive $\alpha$ ($\beta$) biases the propagation of a black (white) cell into the next generation, due to beneficial goods (e.g. an amino acid in short supply by the partner species \cite{MuellerMut}) generated by two nearby cells of the opposite type. Negative $\alpha$ and $\beta$  represent the effect of cells inhibiting the growth of neighboring competing variants.   

 For $d=1+1$, the model is implemented on a square lattice (with one space and one time direction) with periodic boundary conditions in the spatial direction.  During each generation (one lattice row along the spatial direction), the states of all triplets of adjacent cells are used to determine  the state of the middle cell in the next generation using Eq.~(\ref{eq:updaterules}).   For $d=2+1$, we stack triangular lattices of cells (representing successive generations) in a hexagonal close-packed three dimensional array.  Each cell sits on top of a pocket provided by three cells in the previous generation, so Eq.~(\ref{eq:updaterules}) generalizes immediately (see the Appendix for an illustration).

These simple flat front models generate the rich phase diagrams  of Fig.~\ref{fig:PhaseDiagram}$(a)$ and $(b)$.  The  $d=1+1$  diagram in Fig.~\ref{fig:PhaseDiagram}$(a)$  resembles the stepping stone model result  \cite{KorolevMut}.  One feature is a DP2 point, located at $(\alpha ,\beta) = (\alpha_c, \alpha_c) \approx (0.1242,0.1242)$ in our model.    Applying a  symmetry-breaking coefficient $r \equiv \alpha-\beta \neq 0 $ biases the formation of either black $(r>0)$ or white $(r<0)$ cell domains, and the DP2 transition crosses over to DP transitions along a symmetric pair of critical lines $s_c(r) \equiv \alpha_c(r) + \beta_c(r)$ for $r<0$ and $r>0$.      As in typical cross-over phenomena \cite{statmechbook}, the phase boundaries near the DP2 point are given by $ r \sim \pm [s_c (r)-s_c(0)]^{\phi}$, where $s_c(0) =2\alpha_c\approx 0.2484$ and $\phi$ is a cross-over exponent \cite{DRNCrossover}.  We find $\phi \approx 1.9(1)$ (see the Appendix), consistent with  studies of related models  \cite{odor}.  Hence, we confirm  that our model is in the same universality class as Eq.~(\ref{mutlang}).  Thus, many features of our nonequilibrium dynamical models near the transition lines    (e.g., the  power laws governing the phase boundary shape)  will also appear in the various range expansions describable by Eq.~(\ref{mutlang}).

We now study the approach to the DP2 point along the $\alpha = \beta$ line for $\alpha < \alpha_c$.   As   $\alpha$ increases from $-1/3$ to $\alpha_c>0$,   domain boundaries between white and black sectors diffuse more vigorously.    To check that the entire line $-1/3 < \alpha < \alpha_c$ with $\alpha = \beta$  is dominated at long wavelengths by the annihilation of domain wall pairs [see Fig.~\ref{fig:FlatHet}$(a)$],  we study the heterozygosity correlation function $H(x,t)= \langle f(x+y,t)[1-f(y,t)]+f(y,t)[1-f(x+y,t)] \rangle$, where $\langle \ldots \rangle$ is an  ensemble average and an average over points $y$ along the front \cite{KorolevRMP}.  For a random initial condition of black and white cells in equal proportion, $H(x,t)$ can be fit to 
\begin{equation}
H(x,t) = \frac{1}{2}\, \mbox{erf} \left[ \frac{x}{\sqrt{8 D_{\mathrm{eff}} t}} \right], \label{hetcollapse}
\end{equation}
where the fitting parameter $D_{\mathrm{eff}}$ is the effective diffusivity of the domain walls \cite{KorolevRMP}.  The dependence of $D_{\mathrm{eff}}$ on $\alpha$ away from the DP2 point  is consistent with a simple random walk model of domain walls  described in the Appendix, which predicts $D_{\mathrm{eff}} \approx 1/3+\alpha$ [inset of Fig.~\ref{fig:FlatHet}$(b)$].  As we approach the DP2 point ($\alpha \rightarrow \alpha_c^-$) and domain wall branching becomes important, we observe violations of Eq.~(\ref{hetcollapse}), consistent with field theoretic studies \cite{NPRGVoter, CardyTauber}.

When $d=2+1$,  two-dimensional domains at the voter model point $\alpha = \beta = 0$ lack a surface tension and readily ``dissolve'' \cite{KRBBook, dornic}.   Our simulations show that these dynamics allow for a mutualistic phase for all $\alpha = \beta > 0$, with a remarkable pinning of the corner of the  ``wedge'' of mutualism in Fig.~\ref{fig:PhaseDiagram}$(b)$ to the origin.   A similar phenomenon occurs in branching and annihilating random walks, where an active phase exists for any non-zero branching rate for $d=2+1$ \cite{CardyTauber}.  However, our   model is equivalent to the random walk model only for $d=1+1$ \cite{hammal}, and the potential connection in higher dimensions is   subtle. We now describe how we find the shape of the mutualistic wedge for $d=2+1$.

   When $r=0$, we find a voter model transition at $s=s_c(r=0)=0$, and  \textit{any} $s>0$ pushes the system into a mutualistic phase with a non-zero steady-state domain interface density.   A    perturbation $r \neq 0$ pushes the system away from the voter model class by suppressing  interface formation and induces a DP transition at some $s_c(r)>0$.  Upon exploiting cross-over results for a similar perturbation in Ref.~\cite{CDPFT},   we find phase boundaries given by $r \sim {} \pm  s_c(r) /\ln [s_c(r)/s_0]$, where $s_0 \approx 0.551$  is a non-universal constant found by fitting.  The fitting is discussed in more detail in the Appendix. The resulting curves,  plotted in Fig.~\ref{fig:PhaseDiagram}$(b)$, agree well with  simulations.

\begin{figure}[!ht]
\includegraphics[height=2in]{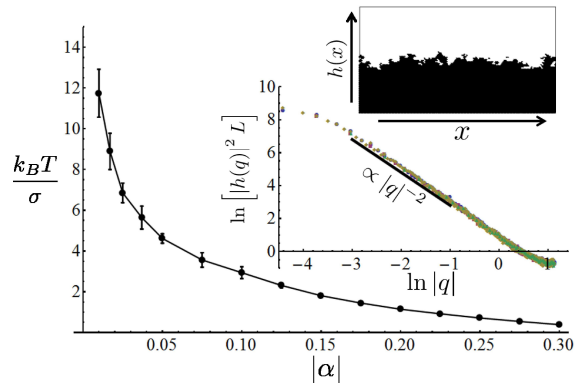}
\caption{\label{fig:CapWave}  (Color online) The main plot shows the effective, dimensionless inverse line tension  $k_B T/\sigma$ in $d=2+1$ dimensions [see Eq.~(\ref{capwave})] for negative $\alpha$ (line guides the eye). Results are from the final interface position $h(x)$ in simulations of an initially flat interface of length $L=512$ evolved for $t \gtrsim7000$ generations.  The interface has overhangs and holes, and $h(x)$ is the average position  for each $x$.  The upper right inset is a sample interface for  $L = 128$ and $t = 500$.       In the remaining inset, we confirm that  Eq.~(\ref{capwave}) correctly predicts the scaling with $L$ by collapsing the Fourier-transformed height $h(q)$, averaged over 160 runs,  for $L = 128,\,256,\,512,\,1024$.  }
\end{figure}

 When $\alpha=\beta<0$ for $d=2+1$,  we find dynamics similar to a kinetic  Ising model   with a  non-conserved order parameter quenched below its critical temperature with an interface density decay $\langle \rho \rangle \sim t^{-1/2}$ \cite{KRBBook} (see the Appendix for more details).      A local ``poisoning'' effect penalizes domain wall deformations, creating an effective line tension $\sigma$ between  domains.    To find $\sigma$, we evolve initially flat interfaces of length $L$  to an approximate  steady-state.   Fluctuations in the interface position $h(x)$ are  characterized by its Fourier transform $h(q)$.   Upon averaging over many realizations, we expect that, in analogy with capillary wave theory \cite{statmechbook}, 
\begin{equation}
\left\langle |h (q)|^2 \right\rangle =   \frac{k_B T}{  \sigma L q^2},
\label{capwave}
\end{equation}
where $T$ is an effective temperature.  Figure~\ref{fig:CapWave} shows that the dimensionless line tension $\sigma/k_BT$  increases as $\alpha$ becomes more negative and that Eq.~(\ref{capwave}) gives the correct prediction for $\left\langle |h (q)|^2 \right\rangle$.  As   we approach the voter model point ($\alpha \rightarrow 0^-$),   $\sigma/k_BT$ vanishes with an apparent power law $\sigma/k_B T \sim |\alpha|^{0.61}$.  However, models with stronger noise might have a voter-model-like coarsening for $\alpha = \beta < 0$, instead  \cite{VoterCrossover}.

 \textit{Rough Front Models.}---We model range expansions with rough frontiers using a modified Eden model  which tracks cells  with at least one empty nearest or next-nearest neighbor lattice site \cite{frey}.  Each such ``active'' cell  $i$ has a birth rate
\begin{equation}
b_i =  \frac{1}{3}+\alpha' N_w(i)  \mbox{\qquad or \qquad} b_i = \frac{1}{3} + \beta 'N_b(i), \label{eq:birthrates}
\end{equation}
if the cell is black or white, respectively.   We set the background birth rates (i.e. for populations of all-black or all-white cells)  to  $1/3$ to make contact with a neutral flat front model. $N_{b,w}(i)$ denote the number of black and white nearest neighbors of cell $i$, respectively.

At each  time step, we pick an active cell $i$ to divide into an adjacent, empty lattice site with probability $b_i/b_{\mathrm{tot}}$,  where $b_{\mathrm{tot}}$ is the sum of the active cell birth rates. The Appendix contains an illustration of this model.   For $d=1+1$, cells can divide into next-nearest neighbor spots to allow for domain boundary branching.   When computing  quantities such as the interface density, we wait for the undulating front to pass and then take straight \textit{cuts}  through the population parallel to the initial inoculation.  The distance of the cuts from the initial inoculation defines our time coordinate.

At the  voter model point  $\alpha' = \beta' = 0$,  the roughness of the front is insensitive to the evolutionary dynamics and genetic domain walls inherit the front fluctuations  \cite{DRNPNAS,Saito}. The average interface density satisfies  $ \langle \rho(t) \rangle\sim t^{-2/3} \sim t^{-1/\tilde{z}}$ \cite{Saito},  where $\tilde{z}=3/2$ represents the  dynamical critical exponent associated with the KPZ equation \cite{KPZ}, or equivalently, the noisy Burgers equation \cite{DRNFS}.   We find that the interface density obeys this scaling for \textit{all} $\alpha' = \beta' < \alpha_c'$  (see the Appendix).
Rough fronts yield novel critical behavior  at the DP2 point  for $d=1+1$:  The cross-over exponent   governing the shape of the phase diagram in Fig.~\ref{fig:PhaseDiagram}$(c)$ decreases considerably to $\phi'  \approx 1.27(15)$, from $\phi \approx 1.9(1)$ for flat fronts. This change leads to a wider mutualistic wedge near the DP2 point.   In addition, in the Appendix, we find a power law decay of the interface density,  $\langle \rho(t) \rangle \sim t^{-\theta_{\mathrm{DP2}}'}$, with a dramatically different critical exponent   $\theta_{\mathrm{DP2}}' \approx 0.61(1)$ compared to $\theta_{\mathrm{DP2}} \approx 0.285(5)$ for flat fronts \cite{NEQPTBook}.     For $d=2+1$, we did not have enough statistics to precisely determine the phase diagram shape.  However, the DP2 point again appears to move to the origin.

\begin{figure}[H]
\includegraphics[height=1.4in]{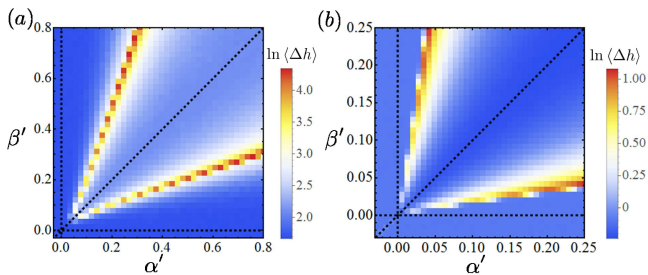}
\caption{\label{fig:roughphase}  (Color online) The fluctuation $\langle \Delta h (t) \rangle$  of the rough frontier (averaged over $10^3$ runs) of   a range expansion  with: (a) $d=1+1$  with $L = 10^3$ cells at time $t=2.5 \times 10^3$ generations;   (b) $d=2+1$ with $L = 50^2$ cells at time $t=500$ generations. The black dotted lines indicate the loci $\alpha'=0$, $\beta'=0$, and $\alpha' = \beta'$.  }
\end{figure}

 The front roughness is a remarkable barometer of the onset of mutualism.  We characterize the roughness by calculating the interface height $h(\mathbf{x},t)$ and its root mean square fluctuation $\langle \Delta h(t) \rangle \equiv \sqrt{\langle ( h(\mathbf{x},t) - \langle h (\mathbf{x},t) \rangle )^2  \rangle }$, where $\langle \ldots \rangle$ is both an ensemble average and an average over all points  $\mathbf{x}$ at the front.  Front fluctuations are greatly enhanced along the pair of DP transition lines   for $d=2+1$ and $d=1+1$, as shown    in Fig.~\ref{fig:roughphase}.   At long times, the roughness  saturates due to the finite system size   (see Ref.~\cite{frey} and the Appendix).

 \textit{Conclusions.}--- To summarize, we found that a mutualistic phase is more accessible in three-dimensional than in two-dimensional range expansions.  Also, antagonistic interactions between genetic variants in three dimensions create an effective line tension between genetic domains.  The line tension vanishes at a neutral point where the variants do not interact and where the mutualistic phase ``wedge'' gets pinned to the origin. In addition to the power laws governing the phase diagram shapes in Fig.~\ref{fig:PhaseDiagram}$(a,b,c)$, we found a striking interface roughness enhancement at the onset of mutualism.  These results  should apply to a wide variety of  expansions because they are insensitive to the microscopic details of our models along transition lines, where we expect universal behavior at large length scales and long times.  The existence of universality has been established for flat fronts \cite{Hinrichsen} and a recent study points to a rough DP universality class  \cite{frey}. 

It would be interesting to compare two- and three-dimensional range expansions  of microorganisms \cite{MuellerMut, WashingtonMut} to test the predicted pinning of the mutualistic phase to the $\alpha = \beta = 0 $ point.  In two dimensions, these expansions are readily realized in Petri dishes \cite{KorolevBac,KorolevMueller}.  In three dimensions, one may, for example, grow yeast cell \textit{pillars}  on a patterned Petri dish with an influx of nutrients at one end of the column, or study the frontier of a growing spherical cluster in soft agar \cite{hersen}.

We thank M. J. I. M\"uller and K. S. Korolev for helpful discussions.  This work was supported in part by the National Science Foundation (NSF) through grants DMR-1005289 and DMR-1306367, and by the Harvard Materials Research Science and Engineering Center via grant DMR-0820484.    Portions of this research were done during a stay at the Kavli Institute for Theoretical Physics at Santa Barbara, supported by the NSF through  grant PHY11-25915.  Computational resources were provided by the  Center for Nanoscale Systems  (CNS), a member of the National Nanotechnology Infrastructure Network (NSF grant ECS-0335765).  CNS is part of Harvard University.

 \appendix
 
 \section{Supplemental Materials \label{supplement}}

\section*{F\lowercase{lat front models and phase boundaries} }

\begin{figure}[ht]
\centering \includegraphics[height=1.25in]{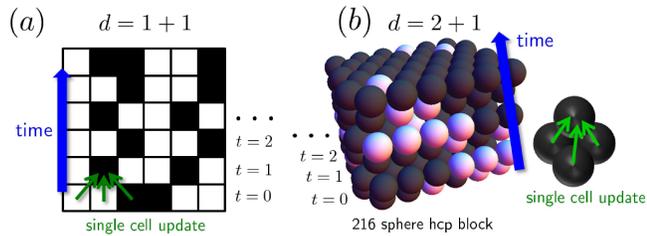}
\caption{\label{fig:flatfrontupdates}  (Color online)  $(a)$ A simulation of a small two-dimensional range expansion with a flat front extending along the horizontal direction.  A single cell update is shown by the three green, thin arrows, with update rules given by Eq.~(2) in the main text.  Each horizontal row of squares in the lattice represents the state of the population front after $t$ generations.  $(b)$ A three-dimensional flat front simulation.  The generations are triangular lattices stacked on top of each other to form a hexagonal close-packed (hcp) array.  A single cell update  in the array is again indicated by the green arrows [probabilities given by Eq.~(2) in main text].    }
\end{figure}

The update rules for the flat front models in two and three dimensions are given by Eq.~(2) in the main text.  As discussed previously, these rules are implemented on a square lattice in two dimensions, and on a close-packed hexagonal array in three dimensions.  Figure~\ref{fig:flatfrontupdates} summarizes the update rules for the two- and three-dimensional models, where one dimension represents time.

\begin{figure}[ht]
\centering \includegraphics[height=2in]{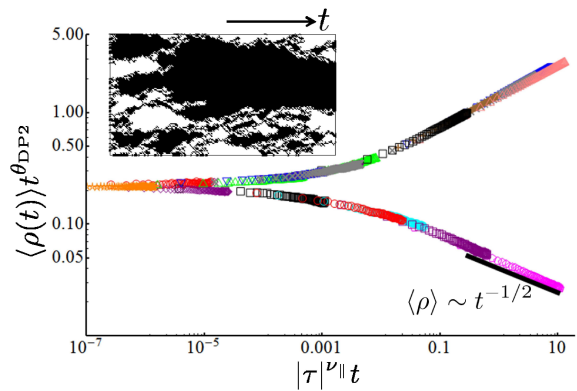}
\caption{\label{fig:flatscaling}  (Color online)  The scaling function for the interface density (averaged over at least 600 runs) near the DP2 transition  ($d=1+1$) along the line $\alpha=\beta$ for  flat front sizes $L \gtrsim 2000$.   Different sets of colored points are different offsets: $\tau = \alpha - \alpha_c=\pm 0.001, \, \pm 0.002 , \ldots$.   The data collapse is consistent with expected DP2 results  $\theta_{\mathrm{DP2}} \approx   0.285(5)$ and  $\nu_{\parallel}  \approx 3.22(6)$ \cite{NEQPTBook}.     The lower branch describes the genetic demixing regime, characterized by the pair-annihilation dynamics of diffusing domain walls. The inset shows the population  for $\alpha = \beta = \alpha_c \approx 0.1242$ at $t=200$ generations and an $L=100$ system size.   }
\end{figure}

In two dimensions we find a demixing regime for parameter values $\alpha = \beta < \alpha_c \approx 0.1242$   as discussed in the main text.  In this regime, we can model the domain walls as pair-annihilating random walkers.    During each generation,  domain walls can  hop to the left or right by one lattice site through updates such as $\circ \circ \circ |  \bullet \bullet \bullet  \rightarrow \circ \circ| \bullet \bullet \bullet \bullet$ (hop to left).   For $\alpha = \beta$, the hops to the left and right occur with equal probability $p$.  If we ignore possible domain branching events in each generation (such as $\circ \circ \circ |  \bullet \bullet \bullet  \rightarrow \circ \circ| \bullet |\circ |\bullet \bullet$), then  $p\approx 1/3+\alpha$.  Hence, we can model the domain walls as random walkers with an effective diffusivity $D_{\mathrm{eff}} \approx p \ell^2/\tau_g \approx 1/3+\alpha$, where $\ell$ is the lattice spacing,  $\tau_g$ the generation time, and we set $\ell = \tau_g = 1$ for convenience.   We can also perform a scaling analysis of the average interface density $\langle \rho(t) \rangle$ for various offsets $\tau = \alpha - \alpha_c$.  We find the scaling function  for $d=1+1$ shown in Fig.~\ref{fig:flatscaling}.   The bottom branch of the scaling function is consistent with the pair-annihilation dynamics in the genetic demixing regime that yields $\langle \rho(t) \rangle \sim t^{-1/2}$.
The data collapse is also consistent with the expected DP2 scaling results for the interface density decay exponent $\theta_{\mathrm{DP2}} \approx 0.285(5) $  and transverse correlation length exponent $\nu_{\parallel} \approx 3.22(6)$  \cite{NEQPTBook}.  

 \begin{figure}[ht]
\centering \includegraphics[height=2.2in]{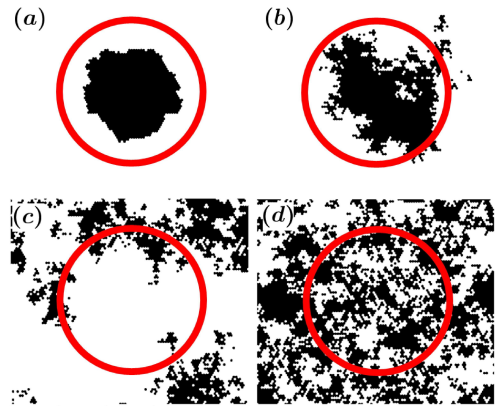}
\caption{\label{fig:droplet}  (Color online) A black droplet with initial radius $R=30$  cells (indicated by thick red circle) is placed in a white cell population.  The droplet is then evolved to $t=5000$ generations using the following parameters: $(a)$ $\alpha = \beta = -0.2$ $(b)$ $\alpha = \beta = -0.02$ $(c)$ $\alpha = \beta = 0$ $(d)$  $\alpha = \beta = 0.02$. }
\end{figure}
 
The dynamics of three dimensional range expansions resembles an overdamped Ising model below $T_c$ with a nonconserved order parameter for $\alpha = \beta < 0$.  The coarsening in the system is driven by a dynamically generated surface tension, calculated in Fig.~3 in the main text using an initially straight, long interface between a black and white cell domain.  Another interesting initial condition is a black cell ``droplet'' in a sea of white cells \cite{dornic}.  Example evolutions for various $\alpha$ and $\beta$ are shown in Fig.~\ref{fig:droplet}.  The dynamical surface tension shrinks the droplet, similar to non-conservative dynamics in an Ising model quenched below the critical temperature.  However, the droplet dissolves away as $\alpha = \beta$ approaches $0$ from negative values and becomes positive.   Interestingly, a related model studied in Ref.~\cite{VoterCrossover}   [based on the discretization of Eq.~(1) in the main text]  exhibits a transition from an Ising model to a voter model coarsening in the presence of strong noise for $\alpha = \beta < 0$.  We did not observe such a transition in our model and the voter model coarsening seems to occur only for $\alpha = \beta = 0$.

\begin{figure}[ht]
\centering \includegraphics[height=2in]{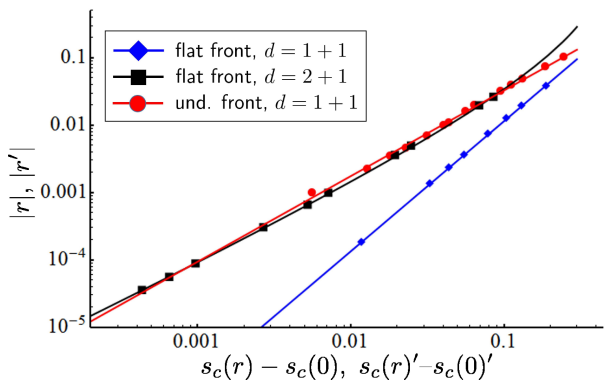}
\caption{\label{fig:phaseboundary}  (Color online) We track the shape of the mutualistic phase boundary by finding the points in the $(r,s)$-plane ($s =\alpha+\beta$ and $r=\alpha-\beta$) for which the average interface density $\langle \rho(t) \rangle$ decays at long times $t$ with a characteristic DP power law $\langle \rho(t) \rangle \sim t^{-\theta_{\mathrm{DP}}}$ (primes for undulating fronts).  We do this for the flat front models for both $d=1+1$  near the special DP2 point (blue diamonds) and for $d=2+1$  dimensions near the voter model point (black squares).  The lines through the points are fits to the functions expected from cross-over scaling: $|r| = A_{\mathrm{2D}} [s_c(r) - s_c(0)]^{\phi}$ for $d=1+1$ and $|r| = A_{\mathrm{3D}} \,s_c(r)/|\ln[s_c(r)/s_0]|$ for $d=2+1$. The fitting parameters are $A_{\mathrm{2D}} \approx 0.970$, $A_{\mathrm{3D}} \approx 0.582$, $s_0 \approx 0.551$, and $\phi \approx 1.93 $.  $A_{\mathrm{2D}}$, $A_{\mathrm{3D}}$, and $s_0$ are non-universal constants and $\phi$ is a cross-over exponent. Similarly, the red circles are points in the $(r',s')$-plane (for the undulating front model in $d=1+1$) near the ``rough DP2'' transition for which $\langle \rho(t) \rangle \sim t^{-\theta'_{\mathrm{DP}}}$, where $\theta_{\mathrm{DP}}' \approx 0.3125$ is the characteristic ``rough DP'' exponent found in Ref.~\cite{frey}.  The red line through the red circles is a fit to $|r'| = A_{\mathrm{2D}}' [s_c'(r') - s_c'(0)]^{\phi'}$.  We find the non-universal amplitude $A_{\mathrm{2D}}' \approx 0.607 $ and the crossover exponent $\phi' \approx 1.27$ by fitting.  }
\end{figure}

To find the shape of the mutualistic ``wedge'' (see Fig.~1 in main text) near the DP2 point for $d=1+1$ and near the voter model point for $d=2+1$, we find the points $(\alpha_i, \beta_i)$ ($i=1,2,\ldots, N$) in the $(\alpha,\beta)$-plane (near the DP2 and voter model points) for which the interface density  $\langle \rho(t) \rangle$ decays at long times $t$ with the expected directed percolation exponent $\langle \rho(t) \rangle \sim t^{-\theta_{\mathrm{DP}}}$,  with $\theta_{\mathrm{DP}} \approx 0.159464$ for $d=1+1$ and $\theta_{\mathrm{DP}} \approx 0.452$ for $d=2+1$ \cite{Hinrichsen}.   In Fig.~\ref{fig:phaseboundary}, we show these $N \approx 10$ points in the $(r,s)$-plane (by transforming to $r_i = \alpha_i - \beta_i$ and $s_i = \alpha_i+ \beta_i$) and fit them to the expected functional forms discussed in the main text.  We find the crossover exponent $\phi \approx 1.9(1)$ by fitting. We estimate the error in $\phi$ by monitoring the variation in the effective exponent $\phi_e(r_i )$     (see, e.g., Ref.~\cite{odor}) for each point $(\alpha_i,\beta_i)$ as we approach the DP2 transition. The exponent is given by
\begin{equation}
\phi_e(r_i)  \equiv  \frac{ \ln [s_c(r_i) -s_c(0)]- \ln [ s_c(r_{i-1})-s_c(0)]}{\ln (r_{i} )- \ln (r_{i-1})}.
\end{equation}
The same analysis can be applied to the rough front model, as shown in Fig.~\ref{fig:phaseboundary}.  Using the effective exponent technique, we find $\phi' \approx 1.27(15)$.

\section*{U\lowercase{ndulating Front Models} }

\begin{figure}[ht]
\centering \includegraphics[height=1.55in]{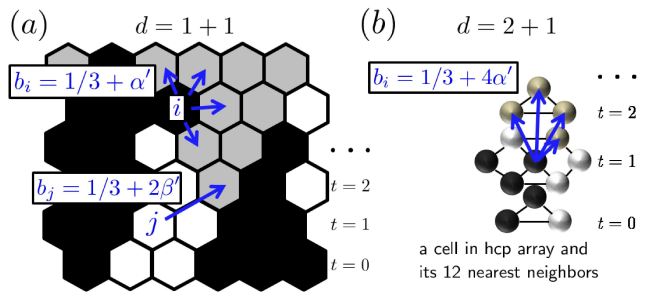}
\caption{\label{fig:roughfrontupdates} (Color online)  $(a)$ A two-dimensional, undulating front simulation is implemented on a triangular lattice of potential cell positions.  The front grows upward, with unoccupied space available to the black and white organisms indicated in grey.  Two examples of the cell growth steps with rates $b_i$ [calculated using Eq.~(5) in the main text] are shown.  Upon selecting a cell to divide, we choose a random empty (grey) nearest neighbor site to place the daughter cell (blue arrows from cell $i$).  If one is unavailable, a random empty next nearest neighbor site is selected (blue arrow from cell $j$).  As discussed in the main text, the time  $t$  (measured in generations) is assigned by taking horizontal cuts through the lattice behind the undulating front.   $(b)$ A three-dimensional, undulating front model is implemented on the  hcp lattice.  We illustrate an update step using a single cell's 12 nearest neighbor sites. The cells have a birthrate $b_i$ [Eq.~(5) in the main text]  calculated using the occupied  nearest neighbor sites (black and white cells)  and daughter cells are placed into empty nearest neighbor sites (four blue arrows into grey spheres).  The background birth rates (without mutualism) are $1/3$ for all cells. }
\end{figure}

\begin{figure}[ht]
\centering \includegraphics[height=1.2in]{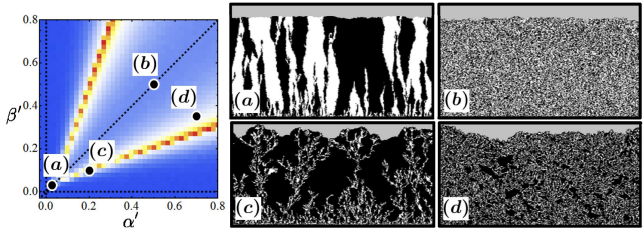}
\caption{\label{fig:SampleRuns}   (Color online) A map of the front roughness at time $t=2500$ generations for an undulating front in $d=1+1$ dimensions with a system size $L=1000$ is shown on the left [see Fig.~4 in main text for the color bar scale].  In $(a)$-$(d)$ we show sample evolutions of the system after about 500 generations at the indicated regions of the map.  The simulation parameters are: $(a)$ The DP2 critical point $\alpha' = \beta' = \alpha_c' = 0.0277$. $(b)$ Symmetric mutualistic phase at $\alpha' = \beta'=0.5$ (c) Near a rough DP transition at $\alpha' = 0.2$ and $\beta'=0.1$. An all black cell front is the absorbing state. (d) An asymmetric mutualistic regime with $\alpha'=0.7$ and $\beta'=0.35$ (black is somewhat favored). }
\end{figure}

It was convenient to implement the undulating front models  on a triangular lattice and on a hexagonal close-packed lattice in two and three dimensions, respectively.  Figure~\ref{fig:roughfrontupdates}  shows how cells are updated in the model.   Note that the interaction strengths $\alpha'$ and $\beta'$ can be negative, but they are bounded from below to ensure that each cell has a positive birthrate [see Eq.~(5) in main text]. We show some sample two dimensional range expansion simulations in Fig.~\ref{fig:SampleRuns} for various interaction strengths $\alpha'$ and $\beta'$.  Note the enhanced roughness near the DP transition line and for antisymmetric mutualism ($\alpha' \neq \beta'$ in the mixed phase) in Fig.~\ref{fig:SampleRuns}$(c)$ and $(d)$.

As discussed in the main text, the dynamics at $\alpha' = \beta'= 0$ for $d=1+1$ is easier to understand as the front roughness is insensitive to the evolutionary dynamics at this point.  The domain interface dynamics inherit the fluctuations of the rough population front and perform super-diffusive random walks.   Figure~\ref{fig:RoughNut}$(a)$ shows an example of the population evolution.    Interestingly, we see in Fig.~\ref{fig:RoughNut}$(b)$  that the scaling function associated with the heterozygosity can be fit to the same error function form as in the flat front case in the demixing regime [Eq.~(3) in main text] if we replace the  scaling variable $x/(8D_{\mathrm{eff}}t)^{1/2}$  by  $x/(Dt)^{2/3}$, where $D$ is an effective ``super-diffusion''  coefficient.

\begin{figure}[ht]
\centering \includegraphics[height=2in]{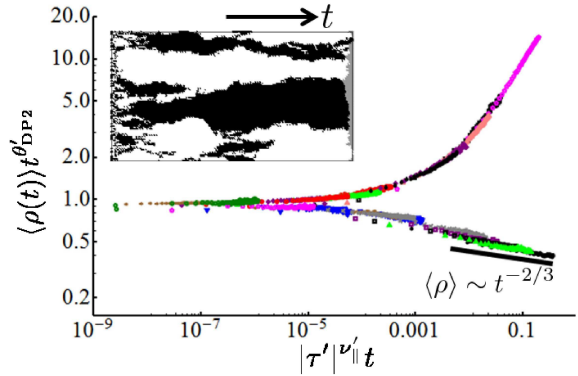}
\caption{\label{fig:roughscaling}  (Color online)   The scaling function for the interface density (averaged over at least 150 runs) near the rough DP2 transition  ($d=1+1$) along the line $\alpha'=\beta'$ for  system sizes $L \gtrsim 2000$.   Different sets of colored points are various offsets: $\tau' = \alpha' - \alpha_c'=\pm 0.001, \, \pm 0.002 , \ldots$.   The data collapse is consistent with the interface density decay exponent   $\theta_{\mathrm{DP2}}' \approx 0.61(1)$ and correlation length exponent $\nu_{\parallel}' \approx 2.8(2) $.    The upper branch of the function describes the mutualistic regime.  The lower branch describes the genetic demixing regime. The inset shows the population  for $\alpha' = \beta' = \alpha'_c$ at $t=200$ generations and an $L=100$ system size.    }
\end{figure}

The scaling function for the average interface density  $\langle \rho(t) \rangle$ along  $\alpha' = \beta'$,  shown in Fig.~\ref{fig:roughscaling},   is consistent with an  exponent   $\theta_{\mathrm{DP2}}' \approx 0.61(1)$  describing the decay of $\langle \rho(t) \rangle$ at the rough DP2 point and a transverse correlation exponent $\nu_{\parallel}' \approx 2.8(2)$.  Note that the flat front results  $\theta_{\mathrm{DP2}} \approx 0.285(5)$ and    $\nu_{\parallel} \approx 3.22(6)$ are markedly different (see  Fig.~\ref{fig:flatscaling} and Ref.~\cite{NEQPTBook}).    Also, the bottom branch of the scaling function in Fig.~\ref{fig:roughscaling}  confirms the rough voter model dynamics,  $ \langle \rho(t) \rangle \sim t^{-2/3}$, in the genetic demixing regime $\alpha'=\beta'<\alpha_c'$.

\begin{figure}[ht]
\centering \includegraphics[height=1.4in]{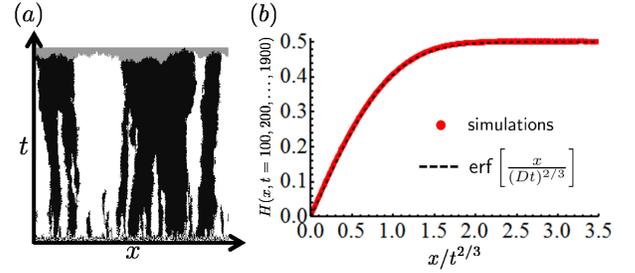}
\caption{\label{fig:RoughNut}  (Color online) (a) Undulating front model simulation in $d=1+1$ dimensions at $\alpha'=\beta'=0$ (system size $L=300$) evolved for about 150 generations. The space-like and time-like axes ($x$ and $t$, respectively) are also shown. (b)  We collapse the heterozygosity $H(x,t)$ for $\alpha'=\beta'=0$ at different times $t=100,200,\ldots,1900$ generations and positions $x$ onto a single universal function using the scaling variable $x/t^{2/3}$, where $2/3=1/\tilde{z}$ is related to the dynamic exponent $\tilde{z}$ of the KPZ model \cite{KPZ, Saito}.   The conjectured analytic form of the scaling function is shown by the black dashed line, where we find  $D \approx 1.1$ by fitting.  The simulation uses an $L=2000$ system size and $H(x,t)$ is averaged over 1600 runs. }
\end{figure}

\section*{T\lowercase{he Rough} DP2 \lowercase{Transition in $d=1+1$ Dimensions}}

The critical exponents along the two rough DP transition lines in $d= 1+1$ dimensions are consistent with those calculated by Frey et al. for a different model exhibiting a rough DP transition \cite{frey}.   The exceptional DP2 transition point where the two lines meet has markedly different scaling properties, however.    Unlike the rough DP transition, the characteristic widths $\xi_{\perp}$ and lengths $\xi_{\parallel}$ of genetic domains (in the space-like and time-like directions, respectively) diverge with different exponents $\xi_{\perp} \sim (\tau')^{-\nu_{\perp}'}$ and $\xi_{\parallel} \sim (\tau')^{-\nu_{\parallel}'}$ as we approach the DP2 point and the offset $\tau'$ approaches zero: $\tau' \equiv \alpha' - \alpha_c' \rightarrow 0$ (with $\tau'>0$).  The exponents $\nu_{\perp,\parallel}'$ are calculated by looking at all compact,   black and white clusters with finite sizes for various points inside the mutualistic phase, i.e., at various  offsets $\tau' \equiv \alpha'-\alpha_c'>0$ from the DP2 point with $\beta' = \alpha'$ .   Following Ref.~\cite{gelimson},
 we calculate
\begin{equation}
\xi_{\parallel} =   \frac{ \sum_i M_i \ell_{\parallel,i}}{\sum_i M_i}   \mbox{\quad and \quad} \xi_{\perp} =   \frac{ \sum_i M_i \ell_{\perp,i}}{\sum_i M_i}   , \label{eq:corrlengths}
\end{equation}
where we sum over all compact clusters $i$.  Each such cluster $i$ contains $M_i$  cells (and hence has ``mass'' $M_i$), and has maximum extensions in the space-like and time-like directions given by $\ell_{\perp,i}$  and $\ell_{\parallel,i}$, respectively.   All lengths are measured in units of the cell diameter.

\begin{figure}[ht]
\centering \includegraphics[height=2.1in]{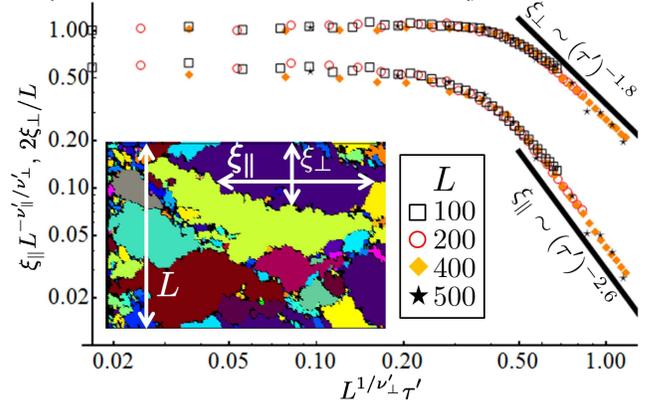}
\caption{\label{fig:CorrLengths}  (Color online) We measure the correlation lengths describing compact domains ($\xi_{\parallel}$ and $\xi_{\perp}$ in inset) using Eq.~(\ref{eq:corrlengths}) for  various front sizes $L$ along the time-like and space-like directions  as we approach the rough DP2 transition at $\alpha' = \beta' = \alpha_c' \approx 0.0277$  for various offsets $\tau' \equiv \alpha'-\alpha_c'>0$ along the line $\alpha'=\beta'$.  We average over at least 150 runs.  The inset shows individual compact domains that we sum over in Eq.~(\ref{eq:corrlengths}) in different colors.  The time-like direction is horizontal.  Our algorithm for determining compact domains of black and white cells excises slivers of mutualistic mixing at black/white boundaries.  This is done using a modified image segmentation algorithm as described in this supplement.   The data for various $L$ are collapsed using finite size scaling \cite{fisherbarber} and we find power law divergences in the lengths $\xi_{\parallel,\perp}$ for large $L$.     }
\end{figure}

  We identify compact clusters in the simulations by using a sequential, two-pass image segmentation algorithm \cite{ComputerVision}.  In the first pass, we scan and label the cells along the space-like direction [left to right in Fig.~\ref{fig:RoughNut}$(a)$] and advance one row at a time along the time-like direction [from bottom to top in Fig.~\ref{fig:RoughNut}$(a)$].     As we scan,  connected regions of cells are partially identified by assigning cells in each region the same label.  We modify the usual segmentation algorithm by  testing each cell for connectivity only if it is the same color as the two cells labelled in the previous row and the two cells to adjacent to it in its own row (taking into account the periodic boundary conditions along the space-like direction).  In the second pass, we complete the segmentation by assigning the same label to all cells that end up part of a connected region after the first pass (see  Ref.~\cite{ComputerVision} for more details).    We then treat all labelled clusters with masses $M_i \leq 3$  or $\ell_{\parallel,i}=1$ as the ``mutualistic phase''  and discard them when performing the summation in Eq.~(\ref{eq:corrlengths}).  The leftover clusters are shown in different colors in the inset of Fig.~\ref{fig:CorrLengths}.  With a few pathological cases at contorted domain walls, the mutualistic phase will be all the cells which  neighbor at least three cells of the opposite type.

After calculating $\xi_{\perp, \parallel}$ for various front sizes $L$ and offsets $\tau'$, we perform a data collapse in  Fig.~\ref{fig:CorrLengths} and find $\nu_{\perp}' = 1.8(2)$ and $\nu_{\parallel}' = 2.6(2)$.  Note that within the error margin, $\nu_{\perp}'$ is the same as the flat front result,  $\nu_{\perp} \approx 1.83(3)$ \cite{Hinrichsen}.  However, the front undulations suppress $\nu_{\parallel}'$ relative to flat fronts (for which  $\nu_{\parallel} \approx 3.22(6) $  \cite{Hinrichsen}). The errors are estimated by varying the exponents and estimating the range of values for which we still find a data collapse.  In the rough DP case studied in Ref.~\cite{frey},  $\nu_{\parallel}' = \nu_{\perp}' \approx 1.6(1)$.  Thus, the clusters have a markedly different, more asymmetric shape near the rough DP2 transition.  This feature makes sense as the more strongly enhanced front roughness along the DP lines [see Fig.~\ref{fig:SampleRuns}$(c)$] results in a stronger coupling between the dynamics in the time-like and space-like directions.   At the DP2 point, this coupling is weaker but still significant: the exponents $\nu_{\perp}'$ and $\nu_{\parallel}'$ are closer to each other than their flat front counterparts.

It is  also possible to study the scaling of the front roughness.  As discussed in Ref.~\cite{frey},  for a fixed population front size $L$ and time $t$, we expect the average interface height fluctuation $\langle \Delta h(t,L) \rangle$ to obey
\begin{equation}
\langle \Delta h(t,L) \rangle \sim \begin{cases}
t^{\gamma} & t \ll t_{\times} \\
L^{\tilde{\alpha}} & t \gg t_{\times}
\end{cases}, \label{eq:heightflucts}
\end{equation} 
where $\gamma$ is the growth exponent, $\tilde{\alpha}$ is the roughness exponent, and the crossover time $t_{\times}$ satisfies $t_{\times} \sim L^{\tilde{z}}$ with dynamic exponent $\tilde{z} = \tilde{\alpha}/\gamma$.  When $\alpha' = \beta' = 0$, we expect the dynamics to fall into the KPZ universality class, for which $\gamma = 1/3$ and $\tilde{\alpha} = 1/2$ (so $\tilde{z}=3/2$). We find consistent results $\gamma \approx 0.31(2)$  and $\tilde{\alpha} \approx 0.48(5)$ for $\alpha' = \beta' = 0$.   We extrapolated the value and error in $\gamma$ at long times by using the effective exponent technique mentioned above. The exponent $\tilde{\alpha}$ was calculated by varying $L$ and measuring the final, saturated roughness at long times.   At the DP2 point, we   find similar results  $\gamma_{\mathrm{DP2}} \approx 0.31(1)$ and $\tilde{a}_{\mathrm{DP2}} \approx 0.48(4)$.  Hence, the dynamics also appear consistent with the KPZ universality class. However, more extensive simulations are necessary to verify that the coupling of the DP2 evolutionary dynamics to the front undulations do not change the KPZ scaling.  Note that the front undulation scaling along the DP phase transition lines is dramatically different from KPZ scaling, as discovered in Ref.~\cite{frey}.  It would also be interesting to check if the interfaces between black and white domains inherit the front fluctuations in the same way as the neutral case at $\alpha'=\beta'=0$, so that the relation $\theta'_{\mathrm{DP2}} = 1/\tilde{z}_{\mathrm{DP2}}$ holds.  As discussed in the main text, we found $\theta'_{\mathrm{DP2}} \approx0.61(1)$.  Hence, if the domain walls also inherit the front undulations, it is possible that the front roughness actually has a slightly larger dynamic exponent than in the KPZ class, with $\tilde{z}_{\mathrm{DP2}} \approx 1.64$ (instead of $\tilde{z} = 3/2$). 

\bibliographystyle{apsrev}
\bibliography{MutBib}

\end{document}